# Insight into the melt processed Polylimonene oxide/Polylactic acid blends

Miguel Palenzuela[a], Juan F. Vega[b,c], Virginia Souza-Egipsy[b], Javier Ramos[b,c], Christian Rentero[b], Sessini Valentina[a]* and Marta E. G. Mosquera[a]*

[a.]Departamento de Química Orgánica y Química Inorgánica, Instituto de Investigación Química "Andrés M. del Río”. Campus Universitario, E-28871 Alcalá de Henares, Spain.

[b.]BIOPHYM, Departamento de Física Macromolecular, Instituto de Estructura de la Materia, IEM-CSIC, c/Serrano 113 bis, 28006 Madrid, Spain.

[c.]Interdisciplinary Platform for Sustainable Plastics towards a Circular Economy, SUSPLAST-CSIC, 28006 Madrid, Spain.

Corresponding Author e-mail: valentina.sessini@uah.es, martaeg.mosquera@uah.es

**ABSTRACT**

In this work, the polymerization of Limonene Oxide (LO) has been optimized at room temperature with two different alluminium-based catalysts [AlMeX{2,6-$(CHPh_2)_2$-4-$^t$Bu-$C_6H_2O$}] (X = Me (**1**), Cl (**2**) ). A fully bio-based ether, polylimonene oxide (PLO), has been synthetized with low molecular weight and good thermal stability, being a potential sustainable polymeric additive for other bio-based and biodegradable polymers such as PLA. Hence, we have explored its ability to influence the thermal, mechanical, and morphological properties of PLA obtaining their blends by melt-processing. The addition of a low amount of PLO led to the decrease of the glass transition temperature of PLA of about 10 °C. Moreover, the decrease of the melting temperature and the degree of crystallinity of PLA was observed. Interestingly, a remarkable increase of the flexibility of PLA based films was noticed. All the results point out the existence of strong interactions between the components suggesting their partial miscibility.

## Introduction

The use of renewable monomers from biomass to produce alternative polymers to traditional oil-derived ones is on increasing demand.[1-8] Among them, monomers coming from non-edible crops and plants arouse particular interest. In this context, terpenes and their derivatives are very interesting substrates as they are produced mainly by plants, and can be found in leaves, flowers, fruits, trees, and spices.[9] Furthermore, terpenes can be produced in biorefineries from biomass coming from different types of waste such as agricultural residues.

Terpenes and terpenoids represent one of the largest families of natural products with a wide structural diversity that can be employed for many applications.[10] From the structural point of view, terpenes contain at least one double bond in their structure and can be considered derivatives of isoprene. These functionalized molecules can be easily derivatized to introduce new functional groups for their use in polymerization and other reactions.[11-13] A particular straightforward modification is the oxidation of the double bond to obtain epoxides, which can be used as monomers in ring opening polymerization (ROP) reactions to produce polymers. Within terpene derivatives, limonene oxide (LO) can be easily produced from limonene, a terpene produced from the peel of some citrus fruits, using conventional epoxidation methods. Although LO has a high potential as monomer due to its multifunctionality, the thermodynamic barrier for the ring opening is high because it is an internal trisubstituted epoxide. As a result, there are limited studies on the polymerization of LO in the literature, with most focusing on copolymerization reactions.[14-38]

LO homopolymerization was firstly reported in 2012 by Park *et al.*[39] as a photoinitiated cationic ring-opening polymerization. However, this method gives numerous side reactions which limited the obtention of high molecular weight polymers. Since then, only our group has reported the synthesis of poly limonene oxide (PLO) using a metal catalyst. We also shown the suitability of the low molecular weight PLO prepared as green biobased additive for polylactic acid (PLA).[40] PLA is an interesting biodegradable polyester that could be a biobased alternative to oil-based polymers. Nevertheless, its brittleness and rigidity have limited its application in many fields.[41] In our previous studied we observed that the addition of only 10 wt % of PLO, led to the improvement of PLA properties in terms of flexibility, thermal stability, and hydrophobicity.[40]

The increasing demand for non-toxic biobased plasticizers fabricated by environmental-friendly strategies[42, 43] have inspired and motivated the current work where we have gone further in our studies and has allowed us to optimize the PLO synthesis as well as the processing method to fabricate PLA/PLO blends using a more sustainable approach. The materials prepared are potential biobased alternatives for food packaging applications and agricultural mulch films. The thermal properties of the neat materials and their blends with different amount of PLO have been characterized as well as their thermal stability, mechanical properties and morphology. Moreover, an in-depth study of the influence of PLO as plasticizer into the PLA behavior has been performed showing their partially miscibility at the right proportions.

## Experimental

### Materials

$AlClMe_2$ (0.9 M in heptane) and (+)-Limonene oxide were purchased from Sigma Aldrich. Poly(lactic acid) (PLA) 2003D, with a density of 1.24 g $cm^{-3}$, a molecular weight ($M_n$) of ca. 1.2 x $10^4$ g $mol^{-1}$, and a melt flow index (MFI) of 6 g 10 $min^{-1}$ (210 °C, 2.16 kg) was supplied by Nature Works®, USA. The aluminium compounds [$AlMeX\{2,6\text{-}(CHPh_2)_2\text{-}4\text{-}^tBu\text{-}C_6H_2O\}$] (X = Me (**1**), Cl (**2**)) were prepared as previously reported by us.[44]

### Synthetic procedures

The monomer was purified by vacuum distillation using $CaH_2$ as drying agent. Once purified, it was stored at −20 °C under argon and in the absence of light. The PLO was synthetized optimizing the procedure previously reported.[40] In particular, the catalyst (0.011 g, 0.02 mmol) was dissolved in the (+)-Limonene oxide (0.818 ml, 5.0 mmol) in the glovebox. The polymerization was done in bulk and inert ambient under magnetic stirring at 25 °C. At the end of the polymerization one aliquot was taken and quenched with wet $CDCl_3$ to determine the conversion of (+)-limonene oxide in polymer by $^1$H-NMR. The conversion in PLO was determined by the integration of the signals observed in the methine region of the monomers at 3 ppm versus the signals of LO isomers above 4 ppm, and the polymer chain centred at 3.5 ppm.

The polymeric samples were purified by washing the remaining monomer and catalyst using methanol and dried at 25 °C. Finally, the sample were dried in a vacuum oven at 60 °C overnight. The final product was characterized by NMR.

**PLA/PLO blends preparation**

PLA and PLO were melt-blended using a microextruder equipped with twin conical corotating screws (MiniLab Haake Rheomex CTW5, Thermo Scientific) with a capacity of 7 $cm^3$. Prior to the melt processing of the samples, the materials were dried in a vacuum oven at 40 °C during 24 h. After weighting the materials, PLA and PLO were manually pre-mixed and then added to the extruder. A screw rotation rate of 50 rpm, temperature of 180 °C, and residence time of 3 min were used. The processed blends were named PLA5PLO, PLA10PLO, PLA15PLO and PLA20PLO highlighting the PLO amount into the blends. The extruded blends were successively thermo-compressed in a Dr. Collin 200mm×200mm press at 180 °C and 100 bars for 5 min, in order to obtain films (≈ 0.5 mm thickness) to carry out their characterization. Neat PLA sample was also prepared following the same methodology for comparison and it was named PLA.

**Characterization methods**

NMR spectra were recorded at 400.13 ($^1H$) and 100.62 ($^{13}C$) MHz on a Bruker AV400 at a room temperature. Chemical shifts (δ) are given in ppm using $C_6D_6$ and $CDCl_3$ as the solvent. $^1H$ and $^{13}C$ resonances were measured relative to solvent peaks considering TMS δ = 0 ppm.

The GC-MS analysis of the LO rearrangement products were performed using a ITQ 900 ion trap Thermoscientific mass spectrometer coupled with a Trace GC Ultra Gas Chromatograph with automatic injector. Electronic impact was used as ionization method within a m/z range of 30-550 Da.

Size-exclusion chromatography analyses were carried out on an Agilent 1260 Infinity II high-speed liquid chromatograph in order to determine the molecular weights ($M_n$ and $M_w$) and polydispersity (Đ) of PLO. Sample solutions (1 mg $ml^{-1}$) in tetrahydrofuran (THF) were injected with a 1 ml $min^{-1}$ flow rate at 35 °C, in two

Mixed D columns connected in series. Calibrations were performed using polystyrene standards.

The thermal stability of the PLO was studied by thermogravimetric analysis (TGA) using a TGA55 analyzer (TA Instrument). Isothermal experiments were performed under oxygen atmosphere to verify its melt-processability at 180 °C. The thermal characterization was performed by dynamic differential scanning calorimetry (DSC). For DSC measurements a DSC 3 Mettler Toledo, Module 444 (Software STARE System SW 14.00 and Intracooler Huber TC45) was used performing a heating/cooling/heating cycles program in the range of 0 to 200 °C with a heating/cooling rate of 10 °C/min and run under nitrogen purge (30 mL/min). The glass transition temperature ($T_g$) was calculated from the second heating scan and was taken at the mid-point of heat capacity changes. The melting temperature ($T_m$), final melting temperature ($T_{end}$) and cold crystallization temperature ($T_{cc}$) were obtained from the second heating, and the degree of crystallinity ($\chi_c$) was determined by using the following Equation:

$$\chi_c = 100 \times \left[\frac{\Delta H_m - \Delta H_{cc}}{\Delta H_m^{100}}\right] \frac{1}{1 - m_f}$$

where $\Delta H_m$ is the enthalpy of fusion, $\Delta H_{cc}$ is the enthalpy of cool crystallization, $\Delta H_m^{100}$ is the enthalpy of fusion of a 100 % crystalline PLA, taken as 93 J/g,[45] and $1- m_f$ is the weight fraction of PLA in the sample. The thermal properties have been also obtained from the first heating run in order to discuss the mechanical properties of the film.

Dynamic Mechanical Analysis (DMA) was performed in a Perkin Elmer DMA7 in the flexural mode at room temperature in the controlled stress mode. Measurements were made in the LVR by dynamic force sweeps between 100 and 3,000 mN at a fixed frequency of 1 Hz. The generated dynamic strain ε* as a consequence of the imposed dynamic stress σ* varied up to a maximum value of 0.10 %, well within the linear viscosity region. The values of the complex flexural modulus *E** at 1 Hz were obtained from the slope of the plot of the imposed applied tensile dynamic stress versus the produced tensile dynamic strain.

Atomic force microscopy (AFM) imaging of the films was carried out using a µTA™ 2990 Micro-Thermal Analyzer (TA Instruments, Inc., New Castle, DE, USA). Topography micrographs were recorded in contact mode at room temperature. A V-shaped silicon nitride probe with a cantilever length of 200 μm and a spring constant of 0.032 N/m was used. The blends were sandwiched between PTFE sheets, heated at 180 °C under minimal pressure for 5 minutes, and isothermally crystallized at 120 °C. The polymeric films have been supported on glass wafers for morphological observations.

## Results and discussion

***Limonene oxide polymerization studies.*** In our previous work, we have assessed the activity of compound [AlClMe{2,6-($CHPh_2$)2-4-tBu-$C_6H_2O$}] (**2**) as catalyst for the ROP of *cis/trans*-(+)-limonene oxide in bulk at 130 °C.[40] In these conditions, PLO chains of low molecular weights were produced within minutes. Interestingly, there are very few catalysts reported in the literature able to reach the high *kinetic activation barrier* that leads to the ROP polymerization of this internal trisubstituted epoxide.[46, 47]

In this study, we have extended our studies to the derivative bearing two methyl groups bonded to the aluminium [$AlMe_2${2,6-($CHPh_2$)2-4-tBu-$C_6H_2O$}] (**1**). As compound **2**, these aluminium derivatives are mononuclear in solution as shown by DOSY experiments.[44]

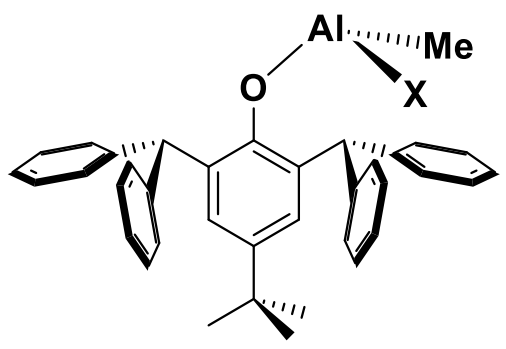

**[AlXMeOAr$^{tBu}$]**

**X = Me (1), Cl (2)**

Figure 1. Aluminium complexes used in this study.

We performed the polymerization in bulk with a catalyst to monomer ratio 1:100 and 1:250 at 130 °C (Scheme 1). As observed for derivative **2** the polymerization took place in only 30 minutes. Furthermore, we explored the influence of the temperature, and we carried out the reaction at room temperature (25 °C), in these conditions both catalysts were also active.

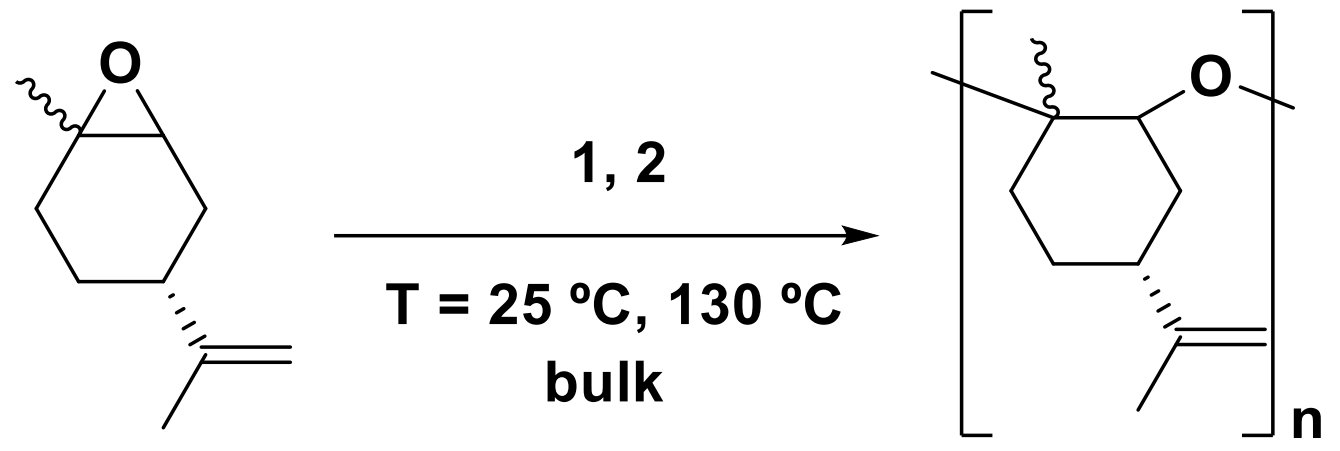

Scheme 1. Catalytic polymerization of (+)-limonene oxide.

The results of the polymerization are reported in Table 1.

Table 1. Experiments of ROP of LO with catalysts **1** and **2**.

| Ent. | [Al] | [Al]:[LO] | T (°C) | %Conv. in PLO [a] | $M_{n,theo}$ (KDa) [b] | $M_{n,GPC}$ (KDa) [c] | PDI [c] |
|---|---|---|---|---|---|---|---|
| 1 | 1 | 1:100 | 130 | 42 | 6.5 | 1.6 | 1.47 |
| 2 | 1 | 1:100 | 25 | 63 | 9.7 | 2 | 1.43 |
| 3 | 1 | 1:250 | 130 | 54 | 20.7 | 2.0 | 1.27 |
| 4 | 1 | 1:250 | 25 | 70 | 26.8 | 2.3 | 1.42 |
| 5 | 2 | 1:100 | 130 | 40 | 6.2 | 1.8 | 1.47 |
| 6 | 2 | 1:100 | 25 | 70 | 10.8 | 1.7 | 1.31 |
| 7 | 2 | 1:250 | 130 | 48 | 18.4 | 1.5 | 1.27 |
| 8 | 2 | 1:250 | 25 | 31 | 11.9 | 2.2 | 1.42 |

[a] Determined by 1H-NMR spectroscopy. b Mn.theo = [LO] / [Al] · %conv. in PLO · 152 + 152. c Determined by GPC-SEC in THF with polystyrene standards.

The activity of both catalysts was similar even though compound **2** was more active in the ROP of other monomers such as glycidilmethacrylate, due to its higher Lewis acid character. [44] In this case the main difference is observed in the molecular weight of the polymers obtained which are slightly higher for the ones prepared using compound **1**, in agreement with a more controlled polymerization. This behaviour has been previously observed for the polymerization of glycidilmethacrylate using these catalysts.[44]

A significant improvement of the reaction was observed when the reaction was carried out at RT. As such, higher conversion into PLO was achieved for both [AlMeX{2,6-$(CHPh_2)_2$-4-$^t$Bu-$C_6H_2O$}] (X = Me (**1**), Cl (**2**)) in comparison to when the reaction was performed at 130 °C (Table 1). The behaviour observed at high temperature could be attributed to the of secondary reactions, such as LO isomerization processes that leads to the formation of side products, which would be more favoured at 130 °C. The increase of the monomer to catalyst ratio led to PLO with similar molecular weight, as shown in Table 1, the obtained polymers showed low molecular weights (~2 kDa), and moderate PDI (~ 1.3). The low molecular weights obtained could be assigned to the occurrence of chain transfer reactions promoted by the alcohol species derived from the rearrangement reactions of the limonene oxide.[40] In fact, we have detected these side products by GC-MS analysis of the LO-polymerization reaction filtrate (see SI, Figures S1-S4). The role of these LO alcoholic derivatives as initiators of the polymerization has been confirmed from the MS analysis of the polymers obtained.[40]

We tried to accomplish a better control over the process by adding a well-known initiator such as benzyl alcohol (BnOH). We performed the reaction of 1 with BnOH and LO with a ratio of 1:1:250, however, the same results as without BnOH were obtained.

We have performed the reaction with the commercial mixture of *cis/trans*-(+)-limonene oxide, and in our initial studies the presence of unreacted *trans*-LO $^{1}$H-NMR spectra made us think that only the *cis*-isomer polymerized.[40] To quantify the real conversion of both monomers, in this work we performed the ROP in the presence of an external standard such as tetraphenylnaphthalene (TPhN). The reaction was carried out with a [Al]:TPhN:LO ratio of 1:2:250. After 30 minutes, 70 % of conversion of the *trans* isomer was observed by $^{1}$H NMR, while total conversion of the *cis* one was achieved, indicating that both isomers polymerize, but *cis*-LO conversion is faster than that of *trans*-LO (see si, Figure S6).

Figures 2 and 3 show the $^{1}$H-NMR and $^{13}$C-NMR spectra, respectively, of the PLO isolated. The broad signals observed in $^{1}$H and $^{13}$C-NMR suggest a random disposition of both monomers in the polymer chain.

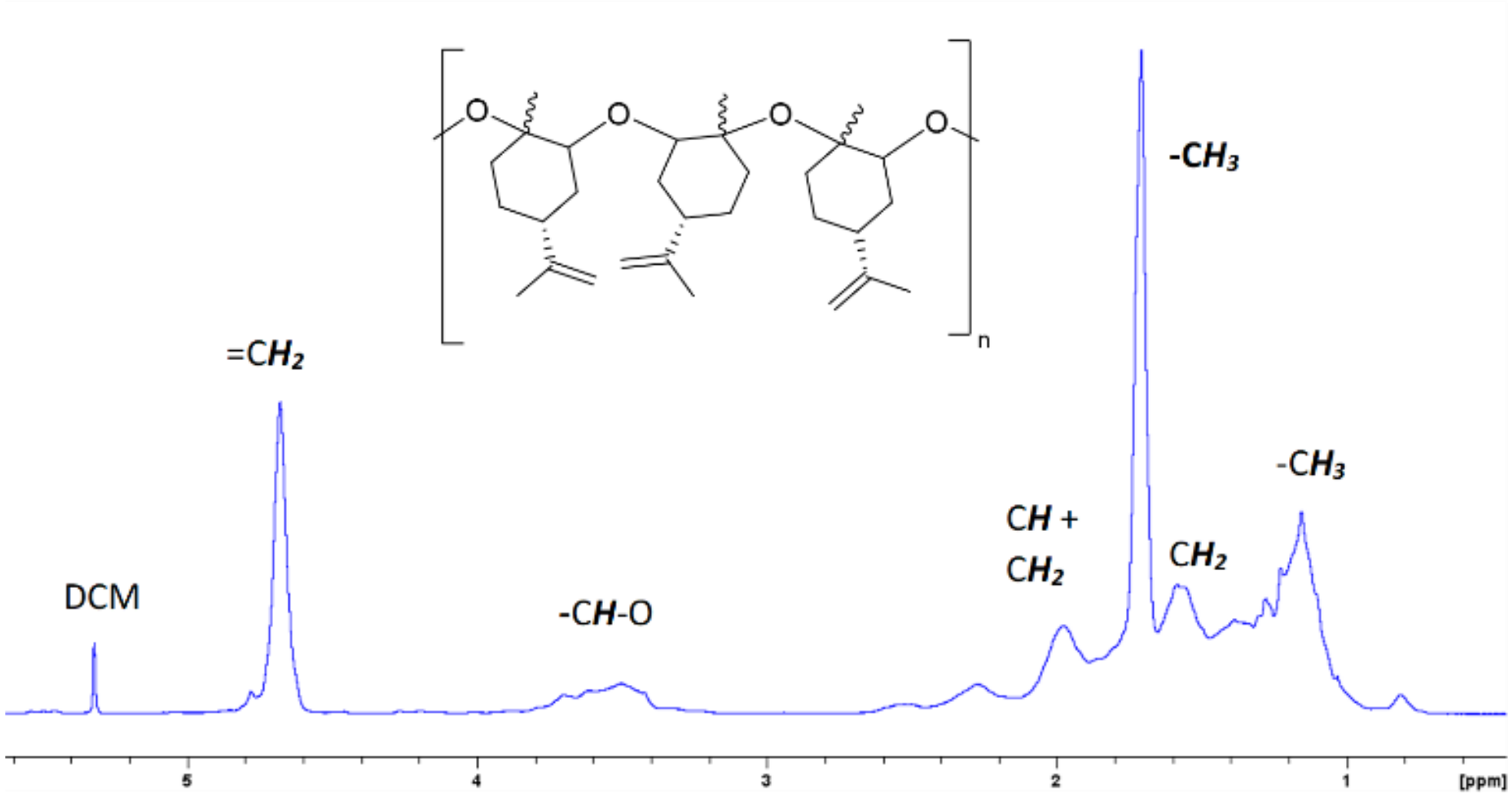

Figure 2. $^{1}$H-NMR spectrum of PLO at RT. $^{1}$H-NMR ($CD_2Cl_2$, 400 MHz, RT): δ (ppm) = 4.68 (s, 2H, *H*2C=C (LO)), 3.20-3.90 (s, 1H, C*H*-O (LO)), 2.69 (m, 1H, C*H* (LO)), 1.96 (m, 1H, C*H*2), 2.17 (m, 1H, C*H*2), 1.97–1.79 (m, 2H, C*H*2), 1.76–1.96 (m, 6H, C*H*3), 1.67 (m, 1H, C*H*2), 1.55 (m, 1H, C*H*2).

As such, for the related poly(limonene)carbonate it has been observed one signal for the *cis* isomer polymer and two for the *trans* one.[18] In the $^{13}$C-NMR of our PLO, at 70-80 ppm appears the signal corresponding to the CH-O carbon which is compose by several peaks in agreement with the polymerization of both isomers.

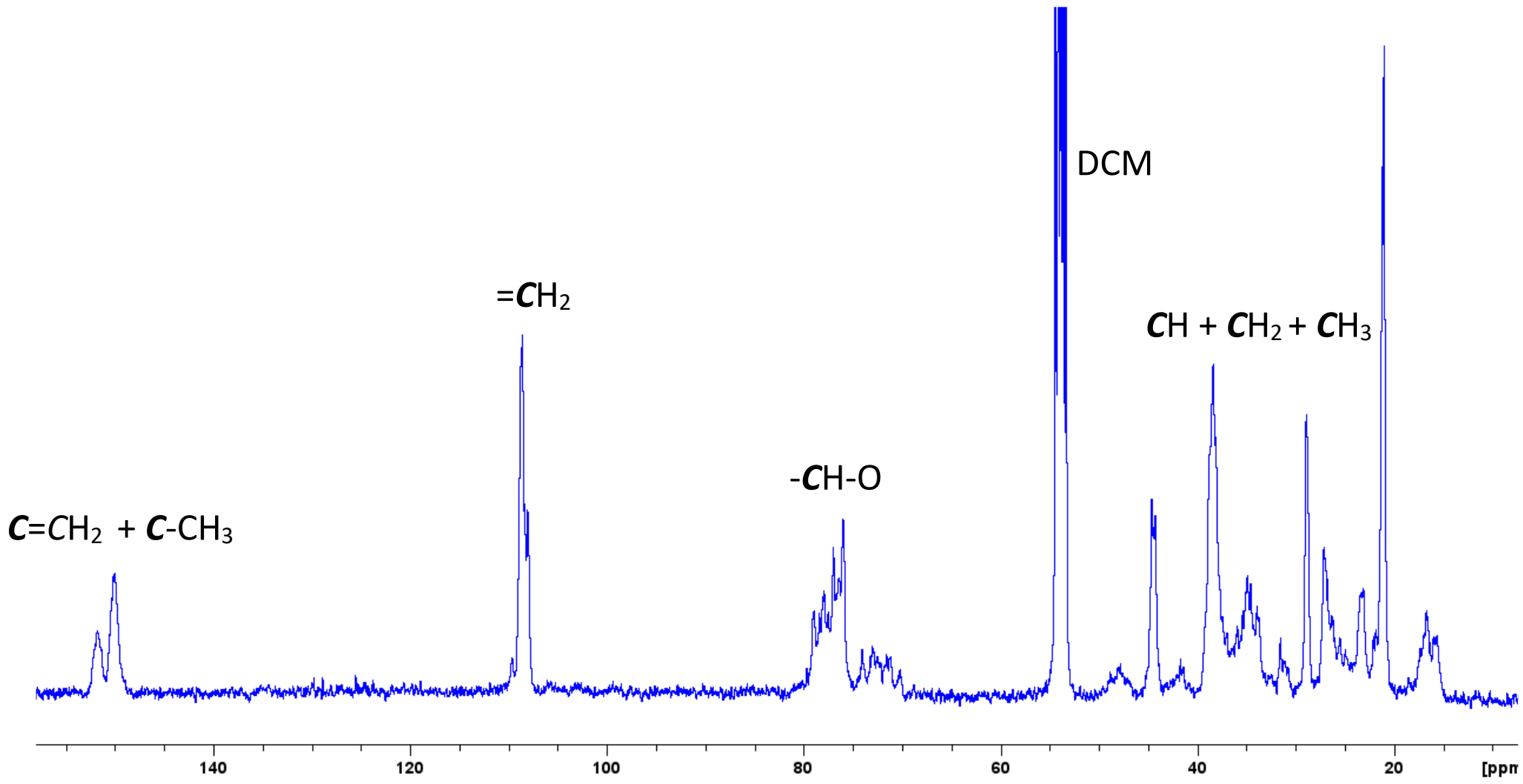

Figure 3. $^{13}$C-NMR spectrum of PLO at RT. $^{13}$C-NMR ($CD_2Cl_2$, 400 MHz, RT): δ (ppm) = 152.0 (Cq, C-CH-O), 150.0 (Cq, C=$CH_2$), 107.4-110.1 (C=$CH_2$), 75.0-80.7 (CH-O), 44.5 (CH), 38.4 ($CH_2$), 34.4 ($CH_2$), 28.9 ($CH_2$), 21.2 ($CH_3$), 16.5 ($CH_3$).

The smaller peaks at higher field corresponds to the polymer terminations (see SI, Figure S7). If both isomers are randomly distributed into the polymeric chain and the epoxide ring is opened every time at the same side, two peaks corresponding to the two different diads, *racemo* and *meso,* would be expected. In our case, the presence of several peaks for each set of signals in the methine region may correspond to different triads and tetrads, indicating that the opening of the epoxide ring has taken place through both sides, giving different conformations. Similar results have been observed by Kleij *et al.*[18] for the copolymerization of $CO_2$ with LO. Therefore, we can propose that an atactic polymer is formed by the polymerization of both isomers of (+)-limonene oxide through the opening on both sides of the epoxide ring (Figure 2).

***PLA/PLO blends studies***. In order to study the effect of PLO on the PLA properties, different amount (5, 10, 15, and 20 wt.%) of PLO have been added to PLA by melt-processing. Firstly, TGA isothermal experiments under air atmosphere were performed on PLO to verify its melt-processability at the processing temperature of PLA, that is 180 °C (see SI, Figure S11). The weight loss was monitored for 15 min at the selected temperature. It is easy to notice that after 3 minutes (residence time used to process the different PLA/PLO blends), PLO weight loss is only 2 % and after 15 minutes is 4 %, indicating that PLO can be melt-blended with PLA without thermal degradation.

The thermal properties of the neat polymers, PLA and PLO, as well as those of their blends have been studied by DSC analysis. The DSC traces ($2^{nd}$ heating runs) obtained for PLA, PLO and their blends are showed in Figure 4 and the thermal properties are reported in Table 2. The DSC thermograms reveal a $T_g$ value for the pure PLA of 57.4 °C. The $T_g$ value for the PLO component is reported to be around 21 °C. In the blends, a shift in the $T_g$ of the PLA component to lower temperatures is reported (see Table 2) indicating a decrease in the rigidity of the polymer matrix due to the incorporation of the low molecular weight PLO plasticizer.

The extent of the $T_g$ depression clearly depends on the concentration, as the decrease is stronger for the highest PLO contents. The degree and the type of interactions between the two components are also factors to consider. The decrease observed in $T_g$ in the blends is typical of miscible and partially miscible blends, in which the plasticizer increases the mobility of polymer chains and reduces the strength of the PLA intermolecular interactions. Prior to the 2nd heating cycle, the samples have been cooled quickly enough, thus a cold crystallization event is observed during heating. This event is observed in all cases, and the temperature peak $T_{cc}$ increases as the amount of PLO does. Finally, at higher temperatures, the melting occurs. In all the samples two well separated peaks are observed, the low-temperature peak that correspond to the melting of the crystalline material present in the samples, and the high-temperature peak associated with the melting of recrystallized material during heating. A clear shift to lower end melting temperatures, $T_{end}$, is observed for all the mixtures. It is interesting to note that the thickening process is still but less visible in the blends, as PLO content increases. For the evaluation of the glass transition and melting depression, the values of $T_g$ and $T_{end}$ have been located and plotted as a function of composition (fraction of second PLO component) in Figure 5a, for the whole set of mixtures studied.

Table 2. Thermal properties (2nd heating run) and mechanical properties of the PLA and PLA/PLO blends

| Samples | $T_g$ (°C) | $T_m$ (°C)[a] | $T_{end}$ (°C) | $T_m$ (°C)[b] | $X_c$ (%)[b] | E (GPa) |
|---|---|---|---|---|---|---|
| PLA | 57.0 | 156.0 | 161.0 | 159.1 | 13.0 | 0.79 |
| PLA5PLO | 55.0 | 155.8 | 160.8 | 159.0 | 7.0 | 0.68 |
| PLA10PLO | 53.0 | 155.3 | 160.3 | 158.4 | 5.4 | 0.60 |
| PLA15PLO | 51.0 | 154.6 | 159.6 | 157.7 | 2.0 | 0.55 |
| PLA20PLO | 50.0 | 154.7 | 158.7 | 156.8 | 1.0 | 0.41 |
| PLO | 21 | - | - | - | - | - |

[a] 2nd peak, [b] 1st run of DSC

We find that both $T_g$ and $T_{end}$ decrease with increasing PLO concentration. The level of the experimental depression is pronounced, much more than the corresponding values of experimental variability in DSC experiments (typically ± 0.1 °C). The level of $T_g$ depression is around 7 °C in the compositional range studied, and it is nicely explained by the Gordon-Taylor[47] and Fox[48] approaches given by Eq. (1) and (2), respectively:

$$T_{g\,blend} = \frac{w_1 T_{g1} + K(1-w_1)T_{g2}}{w_1 + K(1-w_1)} \qquad (1)$$

$$\frac{1}{T_{g\,blend}} = \frac{w_1}{T_{g1}} + \frac{1-w_1}{T_{g2}} \qquad (2)$$

*K in Eq. 1* is $\rho_1\Delta\alpha_2/(\rho_2\Delta\alpha_1)$, being $\rho_1$ and $\rho_2$ the density of the components and $\Delta\alpha_1$ and $\Delta\alpha 2$ the step change in the thermal-expansion coefficient at $T_g$. If we consider a similar value of the density for both PLO (1) and PLA (2) and $\Delta\alpha_i T_{gi}$ a universal constant for polymers,[49] thus $K \cong T_{g1}/T_{g2} \sim 0.9$ such that in the PLA/PLO blends, Gordon-Taylor equation is reduced to the Fox approach given by Eq. 2. The linear dependence of the $T_g$-composition curves in the series studied indicates that weak specific interactions exist between the two components. These weak interactions stablished between PLA and amorphous PLO are also related to the melting depression.

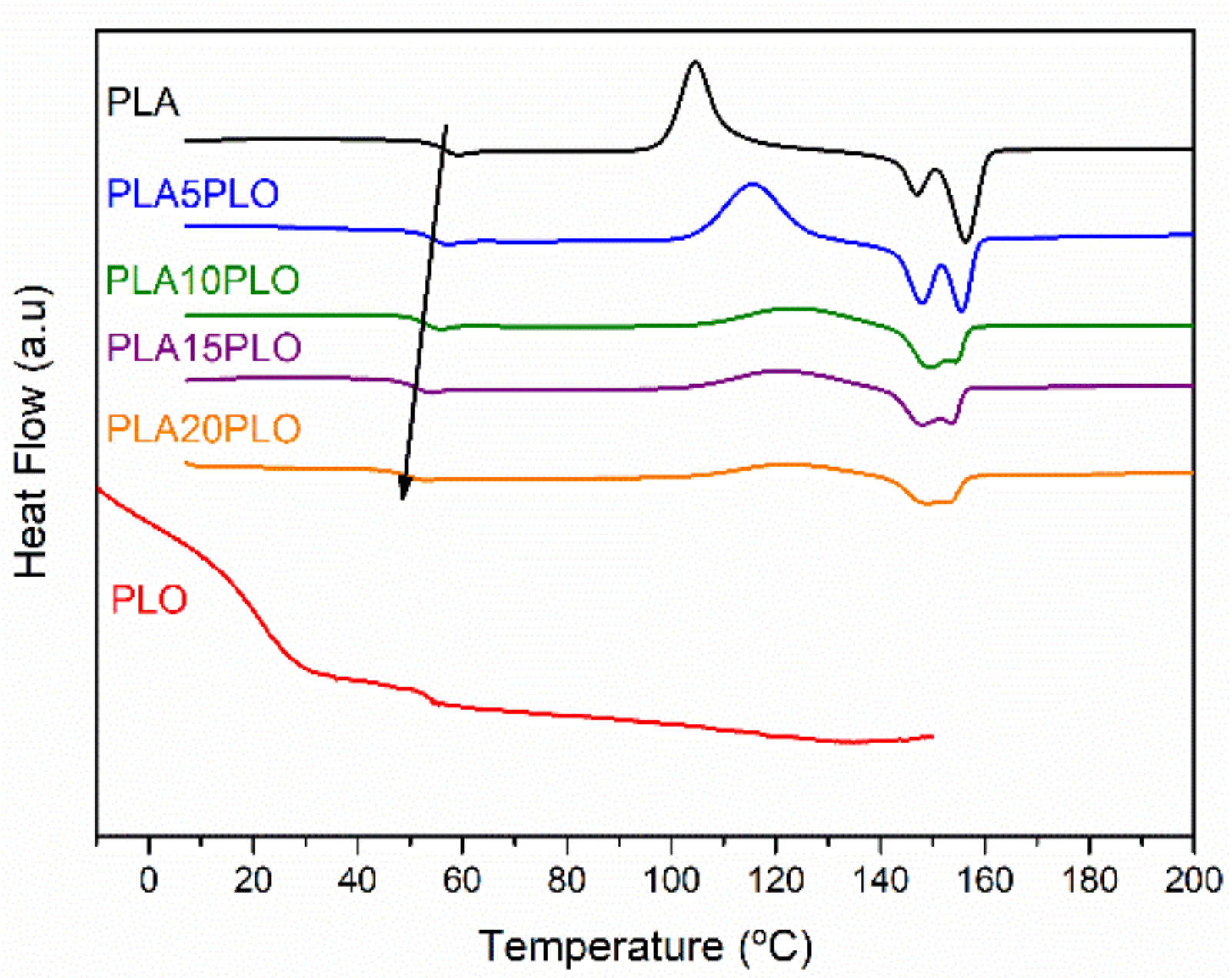

Figure 4. Second heating scan for all the samples

According to Nishi and Wang[50] the thermodynamics of mixing amorphous/crystalline polymers can result in a melting-point depression that can be expressed in relation to the Flory-Huggins interaction parameter, $\chi_{12}$, such that:

$$\Delta T_m = -RT\frac{V_2 T_m^0 \chi_{12}}{V_1 \Delta H_2^0}\phi^2 \qquad (3)$$

R the gas constant $V_1$ and $V_2$ are the molar volumes of the repeating unit of the PLO and the polymer PLA, respectively, $T_m^0$ is the equilibrium melting temperature of PLA, $\Delta H^0_2$ is the enthalpy of fusion per mole of repeating unit of PLA, and $\phi$ is the volume fraction of PLO. The Nishi-Wang equation clearly states a linear dependency of $\Delta T_m$ with $\phi^2$. Of course, the melting temperature changes can be explained not only by the exothermic interaction between the PLA crystal and the amorphous PLO component, but also as due to morphological and kinetic effects as a consequence of the thermal history or because the crystals actually grow far from the equilibrium. The thermal history in the blends studied in this case are assumed to be the same, as in the DSC experiments the memory has been erased during the first heating, and the crystallization step is exactly the same in all cases. Nonetheless, it is clear from the results in Fig. 5b that the linear correlation predicted by Nishi−Wang holds for the system under study. Conventionally, the application of the Nishi−Wang approach requires the determination of the equilibrium melting temperatures provided by Hoffman−Weeks plots.[51] However, PLA and PLO are prone to thermal degradation at crystallization temperatures close to $T_m$, making the use of this procedure problematic for long crystallization times. Therefore, an estimate of the value of $\chi_{12}$ was performed as suggested by Pizzoli et al.[52] using the nonequilibrium melting peak and end temperatures determined by DSC instead (see Table 2). Considering Eq. 3 and the linear fit in Figure 5b, the value for $\chi_{12}$ can be obtained from the slope. Considering the values $V_1$=141.3 mL/mol, $V_2$ = 59.3 mL/mol (PLA), $\Delta H$ = 6,752 J/mol, the interaction parameter is found to be small and negative ($\chi_{12}$ = -0.6), indicating that PLA and PLO are thermodynamically miscible in the melt, or at least that interact likely due to hydrogen bonding.

We have additionally evaluated the mechanical properties of the films prepared from compression moulding. Obviously, the mechanical properties of such samples will depend on their initial microstructure and morphology. The details of this microstructure may be understood from the DSC features obtained in the first melting cycle, as listed in Table 2. A decrease of the crystal content as PLO content increases is envisaged from the data, pointing towards again to a plasticization effect of PLO. This fact directly affects to the mechanical properties as observed in Figure 6. The dependence of the dynamic bending stress, $\sigma_b$*, with the dynamic strain, $\varepsilon_b$*, (dynamic stress-strain curves) at a frequency of 1 Hz in the samples are observed in Figure 6a. The results represent the average obtained from three independent measures in each sample.

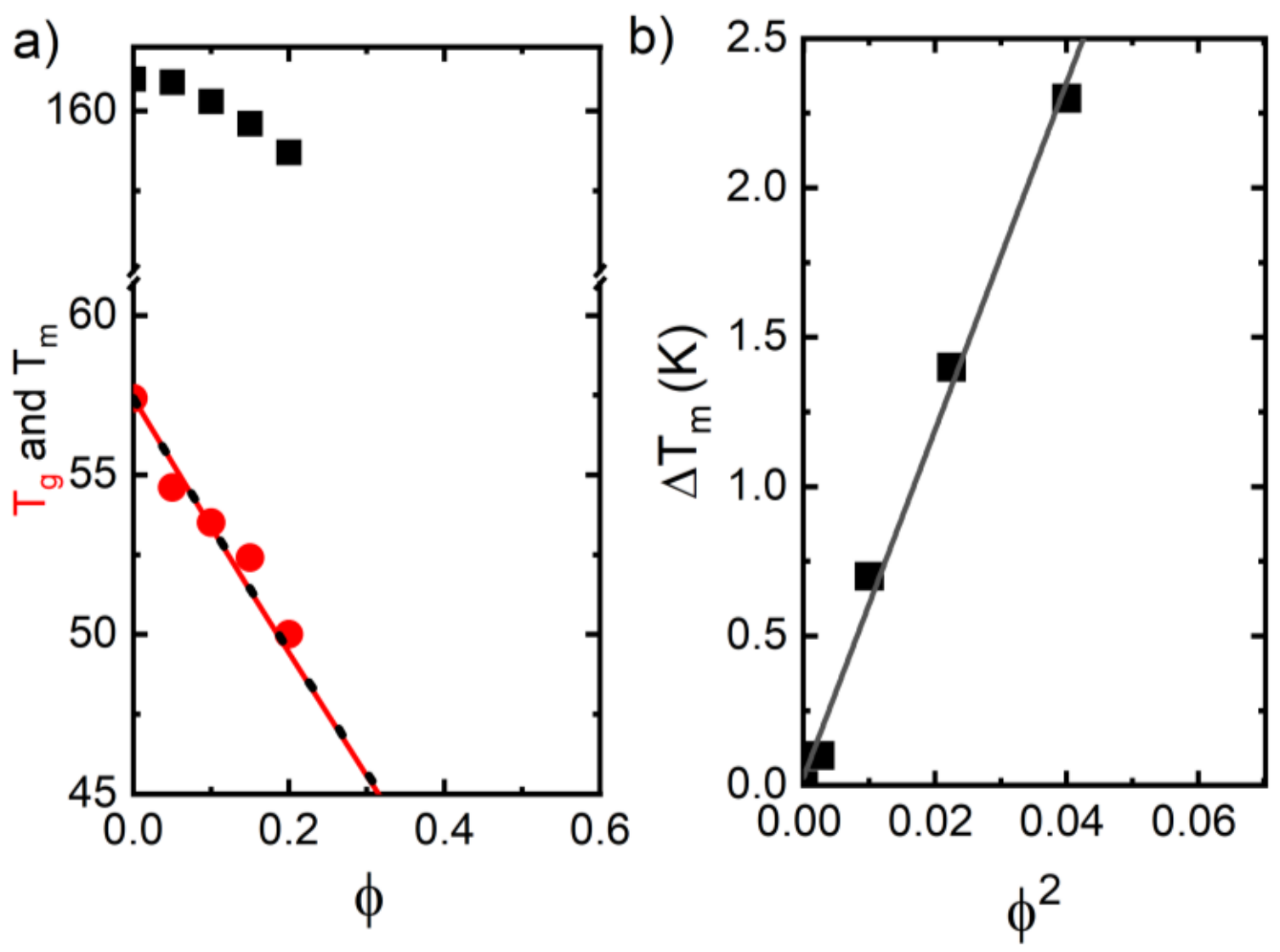

Figure 5. A) $T_g$ (circles) and $T_{end}$ (squares) compositional dependence in the samples under study. The solid line represents the fit to the Fox equation, and the dashed line the corresponding fit to the Gordon-Taylor approach. B) Melting temperature depression obtained for the same set of samples. The line represents a linear fit to the Nishi-Wang approach.

The dynamic flexural stress–strain curves show a large difference between the pure PLA sample and the blends. The values of the slope for small values of the dynamic strain (linear region) directly give to the dynamic flexural modulus, E*.

PLA shows a value of the flexural modulus of 0.80 GPa. PLA/PLO samples show the typical behaviour associated to plasticisation, as the tensile modulus decreases with the presence of PLO to a value of around 0.40 GPa for the blend with the highest PLO content (Figure 6b). Low-$M_w$ plasticizer actually behaves as solvent molecules, leading to a decrease of the density of interactions among PLA macromolecules. These results again point towards a certain level of miscibility between PLA and PLO.

The morphology of neat PLA was compared to that of PLA blended with 10 % PLO. The morphological aspects are observed in Figure 7 for films prepared by complete crystallization at T = 120 °C from the melt.

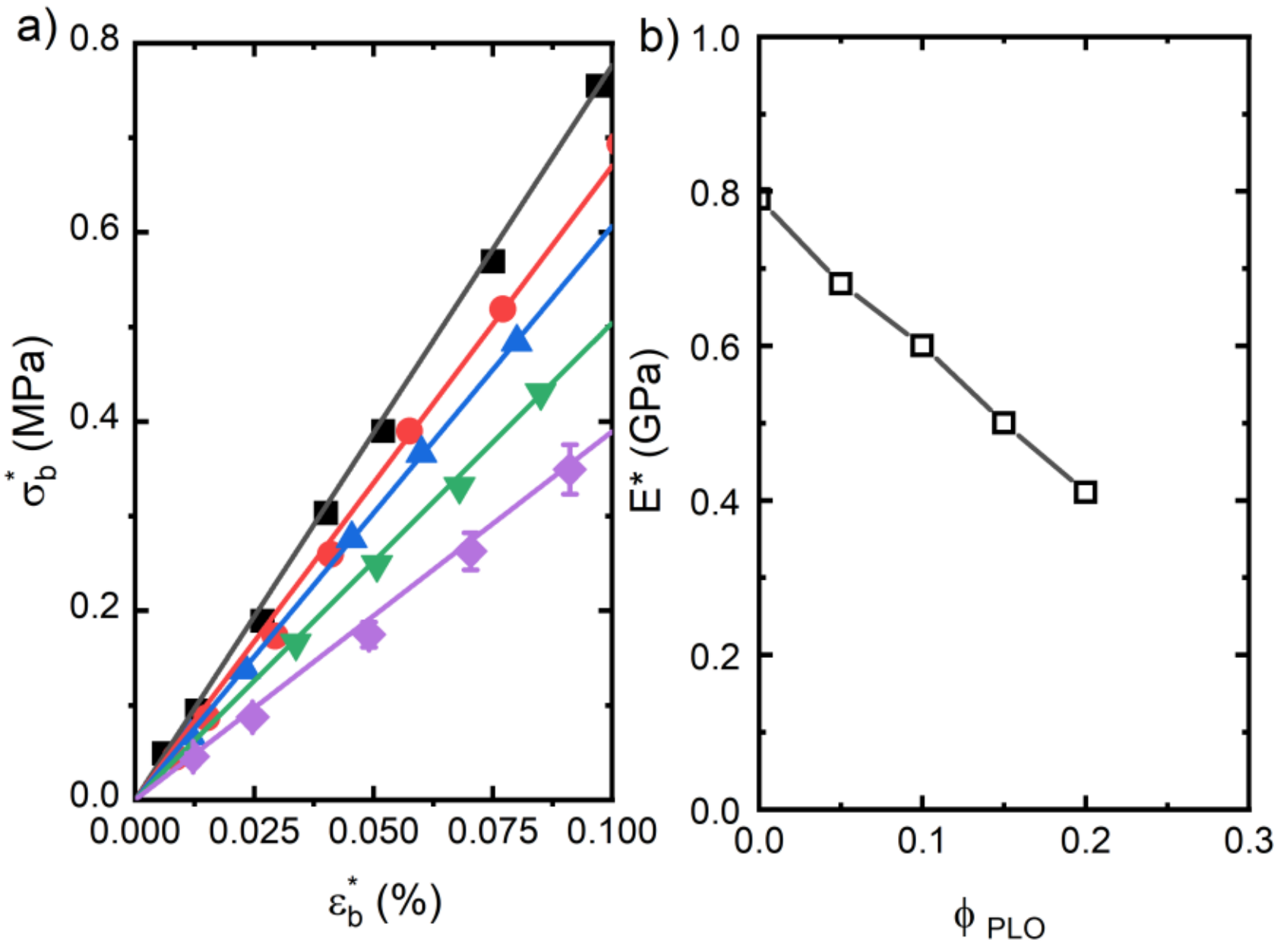

Figure 6. (A) Dynamic bending stress - strain curves at 1 Hz and room temperature of the samples studied, (■) PLA, (●) PLA5PLO, (▲) PLA10PLO, (▼) PLA15PLO and (◆) PLA20PLO. The lines are the linear fits to experimental data. (B) Flexural modulus compositional dependence in the samples under study at T = 20 °C.

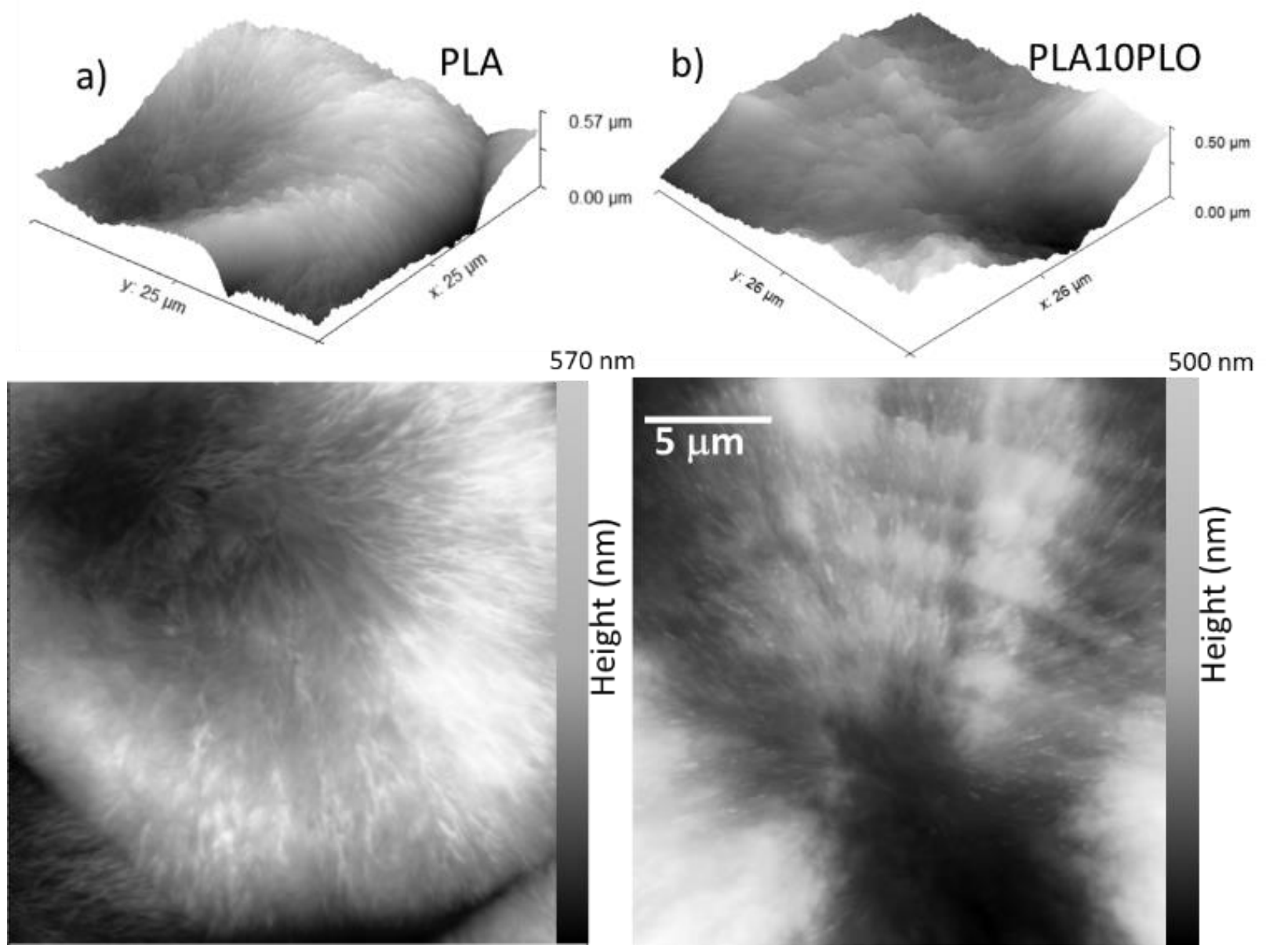

Figure 7. AFM image of the spherulitic morphology of PLA (left), (B) PLA10PLO (right).

The AFM images in Figure 7 indicate the absence of a ring-band pattern in the neat PLA (left panel). However, the presence of PLO alters the spherulitic morphology, as observed in the right panel B. The AFM images shows that the distance between ridges is approximately 2 μm, and the height difference between the ridges and valleys is around 20 - 50 nm. Furthermore, irregular intermediate ridges are present between the long ridges and the valleys. This hierarchical organization of the crystalline structure within the ridges and valleys suggests some type of segmental interaction and partial miscibility between the PLA and PLO. The growth of spherulites with ringed bands is a commonly reported phenomenon in blends with interactions between the components. This phenomenon has been particularly observed in polyesters mixed with polar diluents, as documented in a study by Keith et al.[53] Based on these findings, the morphological characteristics observed in the PLA/PLO blends under study may be attributed to segmental interactions and partial miscibility between the two components.

## Conclusions

The aluminium catalysts [AlMeX{2,6-$(CHPh_2)_2$-4-$^t$Bu-$C_6H_2O$}] (X = Me (**1**), Cl (**2**)) are active catalysts for the ROP polymerization of LO even at room temperature, leading to greener synthetic parameters. This allowed us to obtain poly limonene oxide totally derived from renewable sources under sustainable conditions. The polyether obtained had low molecular weight and good thermal properties being a promising green additive to be melt-blended with other bioplastics such as PLA.

The PLA/PLO blends show a decrease in the glass transition and melting point of the main PLA component. In fact, the application of Fox and Nishi-Wang approaches points to the existence of interactions between the components suggesting partial miscibility. The mechanical properties of the blends also showed plasticization, with decreased elastic modulus resulting in less fragile systems compared to neat PLA.

These sustainable materials are entirely derived from renewable sources and their obtention is fully scalable by industrial methods used for traditional polymers, being interesting for many applications such as biodegradable food packaging or agricultural mulch films.

**Author Contributions**

The manuscript was written through contributions of all authors. All authors have given approval to the final version of the manuscript.

**Conflicts of interest**

There are no conflicts to declare.

**Acknowledgements**

The authors would like to thank the financial support from the Comunidad de Madrid (EPU-INV/2020/001) and the Ministerio de Ciencia e Innovación (Spain) through the projects TED2021-130871B-C22, PID2021-122708OB-C31, PID2019-107710 GB-I00, and RYC2021-033921-I. BIOPHYM Service at the IEM-CSIC is acknowledged for granting the use of the facilities. This project has received funding from the European Union's Horizon 2020 research and innovation programme under the Marie Skłodowska-Curie grant agreement No 754382, GOT ENERGY TALENT. The content of this article does not reflect the official opinion of the European Union. Responsibility for the information and views expressed herein lies entirely with the authors.

# Supporting Information

# Insight into the melt processed Polylimonene oxide/Polylactic acid blends

Miguel Palenzuela[a], Juan F. Vega[b,c], Virginia Souza-Egipsy[b], Javier Ramos[b,c], Christian Rentero[b], Sessini Valentina[a]* and Marta E. G. Mosquera[a]*

[a.]Departamento de Química Orgánica y Química Inorgánica, Instituto de Investigación Química "Andrés M. del Río". Campus Universitario, E-28871 Alcalá de Henares, Spain.

[b.]BIOPHYM, Departamento de Física Macromolecular, Instituto de Estructura de la Materia, IEM-CSIC, c/Serrano 113 bis, 28006 Madrid, Spain.

[c.]Interdisciplinary Platform for Sustainable Plastics towards a Circular Economy, SUSPLAST-CSIC, 28006 Madrid, Spain.

Corresponding Author e-mail: valentina.sessini@uah.es, martaeg.mosquera@uah.es

CONTENTS

# 1. Characterization of LO rearrangement products by GC-MS

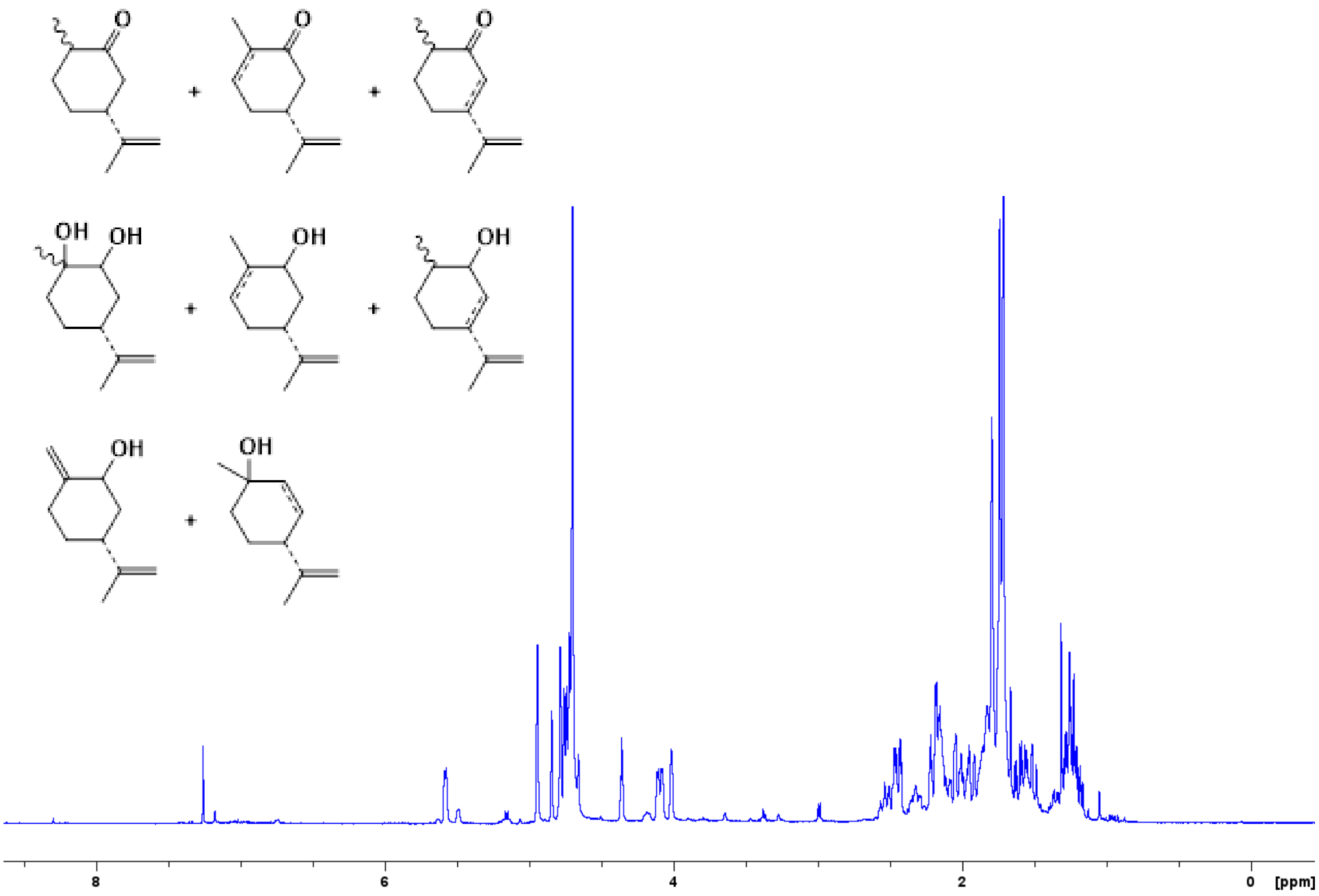

**Figure S1.** $^{1}H$-NMR spectrum of the filtrate of the ROP reaction of LO with catalysts **1** where LO side products cannot be identified.

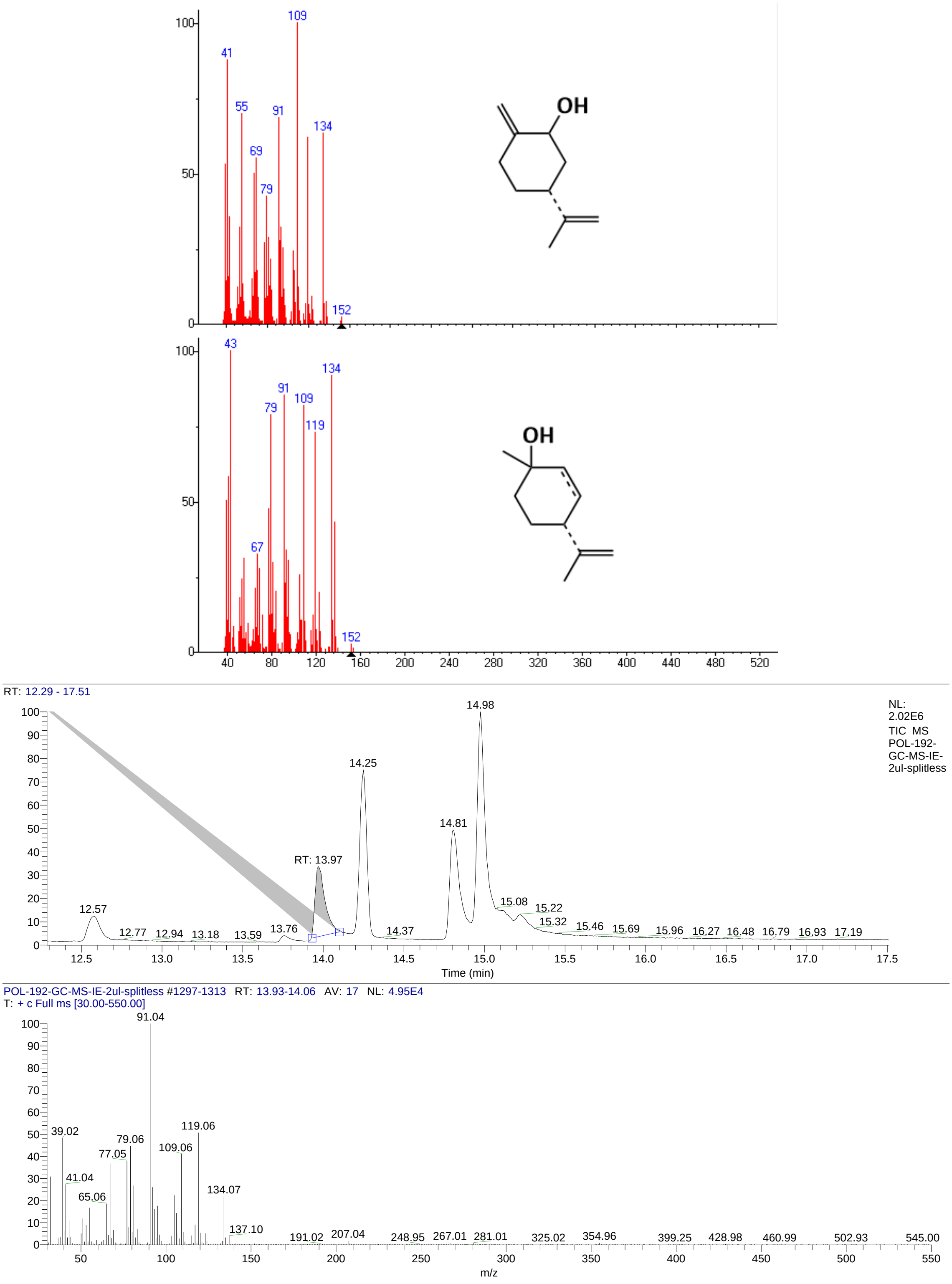

**Figure S2.** GC-MS of LO alcohol derivatives present in the filtrate of the ROP reaction of LO with catalysts **1.**

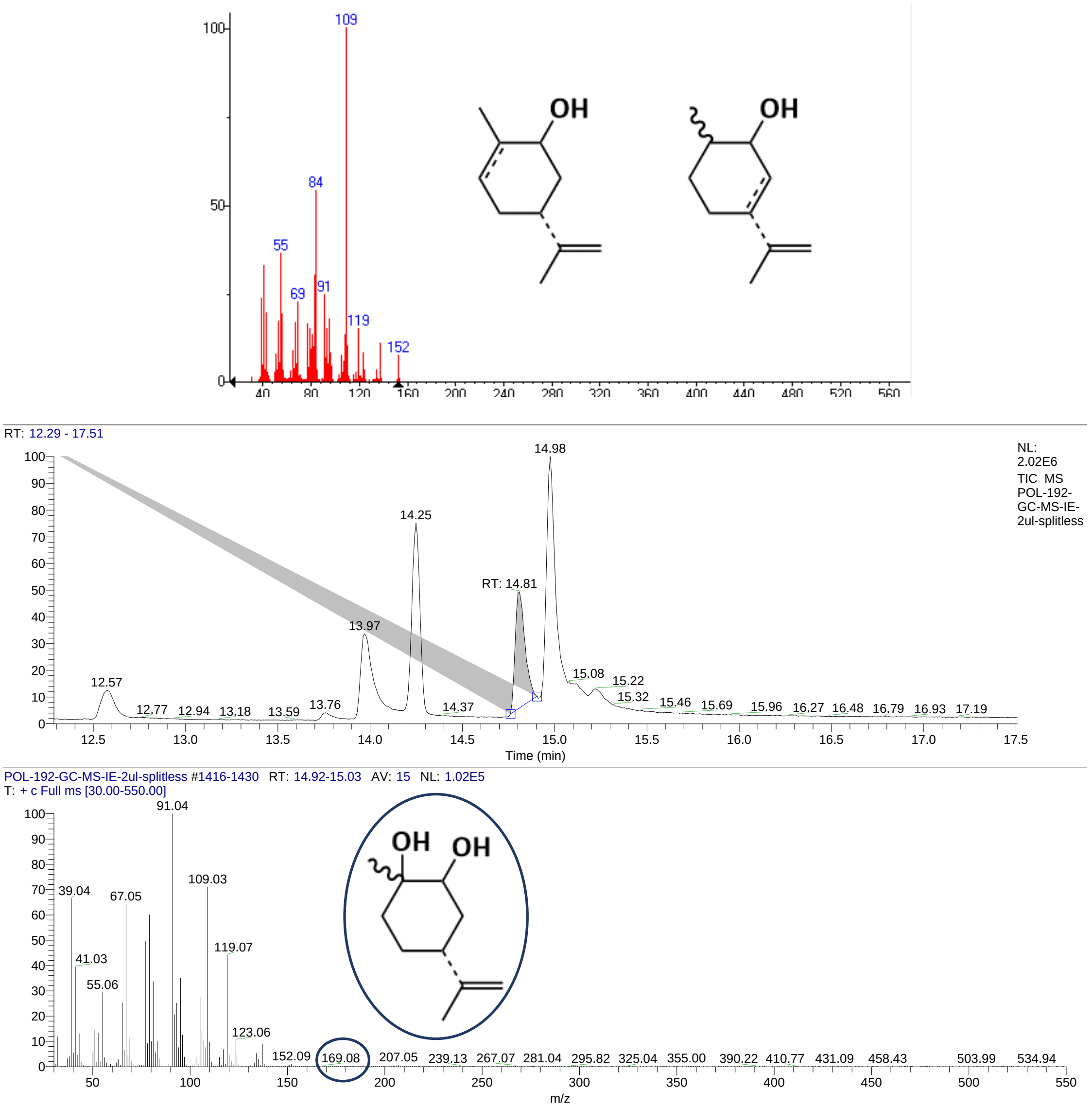

**Figure S3.** GC-MS of LO alcohol derivatives present in the filtrate of the ROP reaction of LO with catalysts **1.**

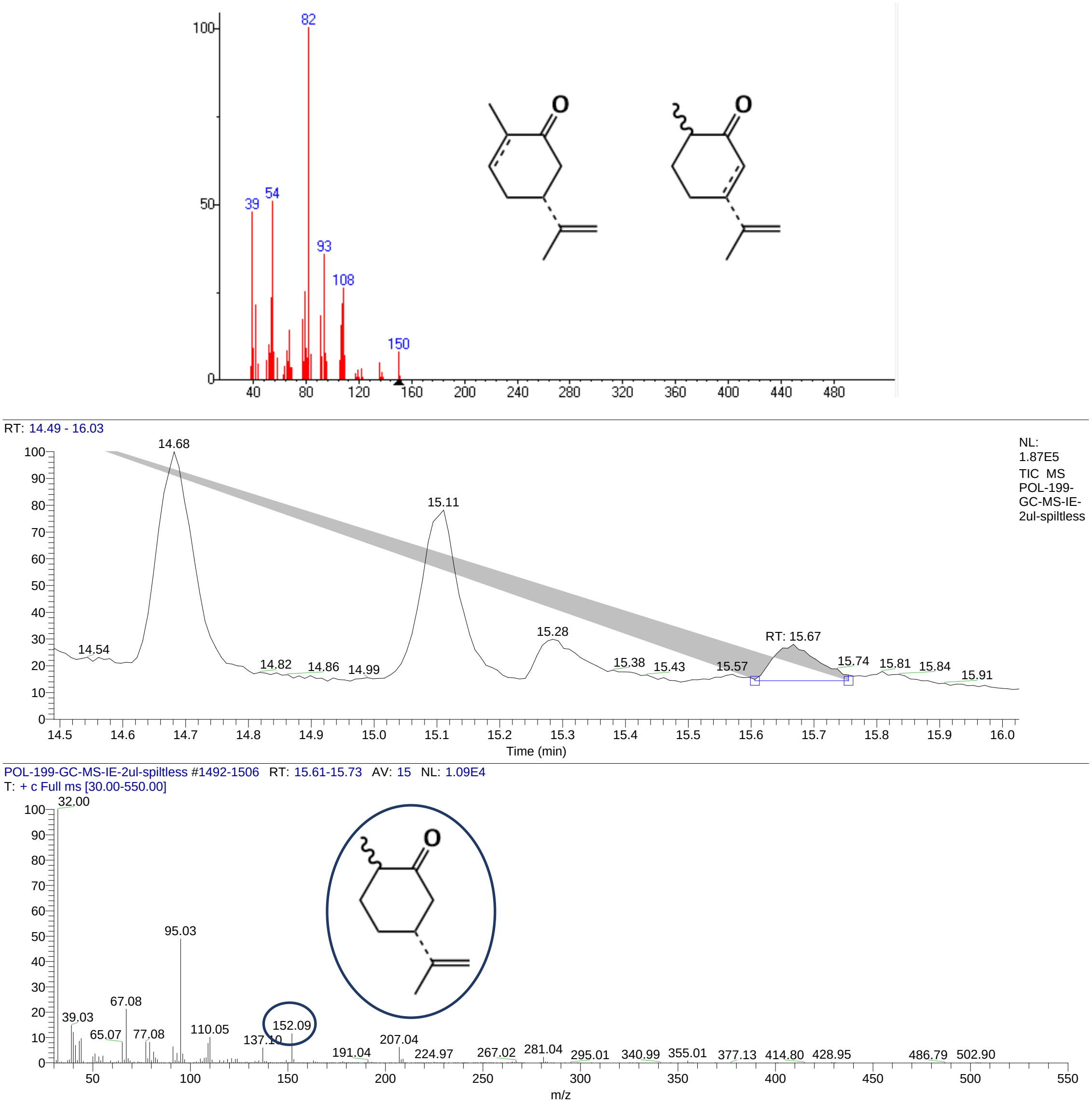

**Figure S4.** GC-MS of LO ketone derivatives present in the filtrate of the ROP reaction of LO with catalysts **1.**

## 2. NMR spectra

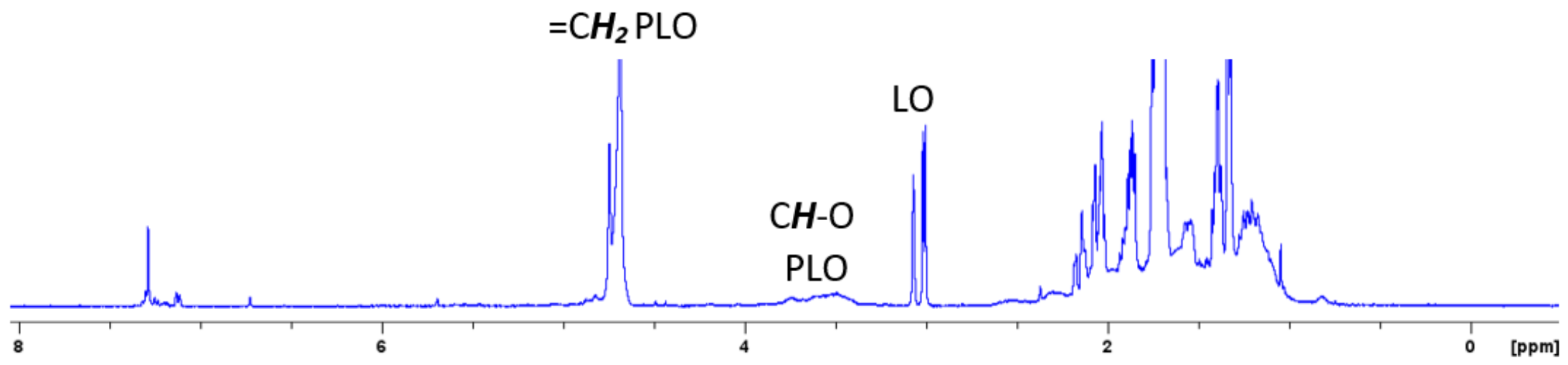

**Figure S5.** [1]H-NMR spectrum of an aliquot of the reaction of 250 equivalents of LO in bulk at RT.

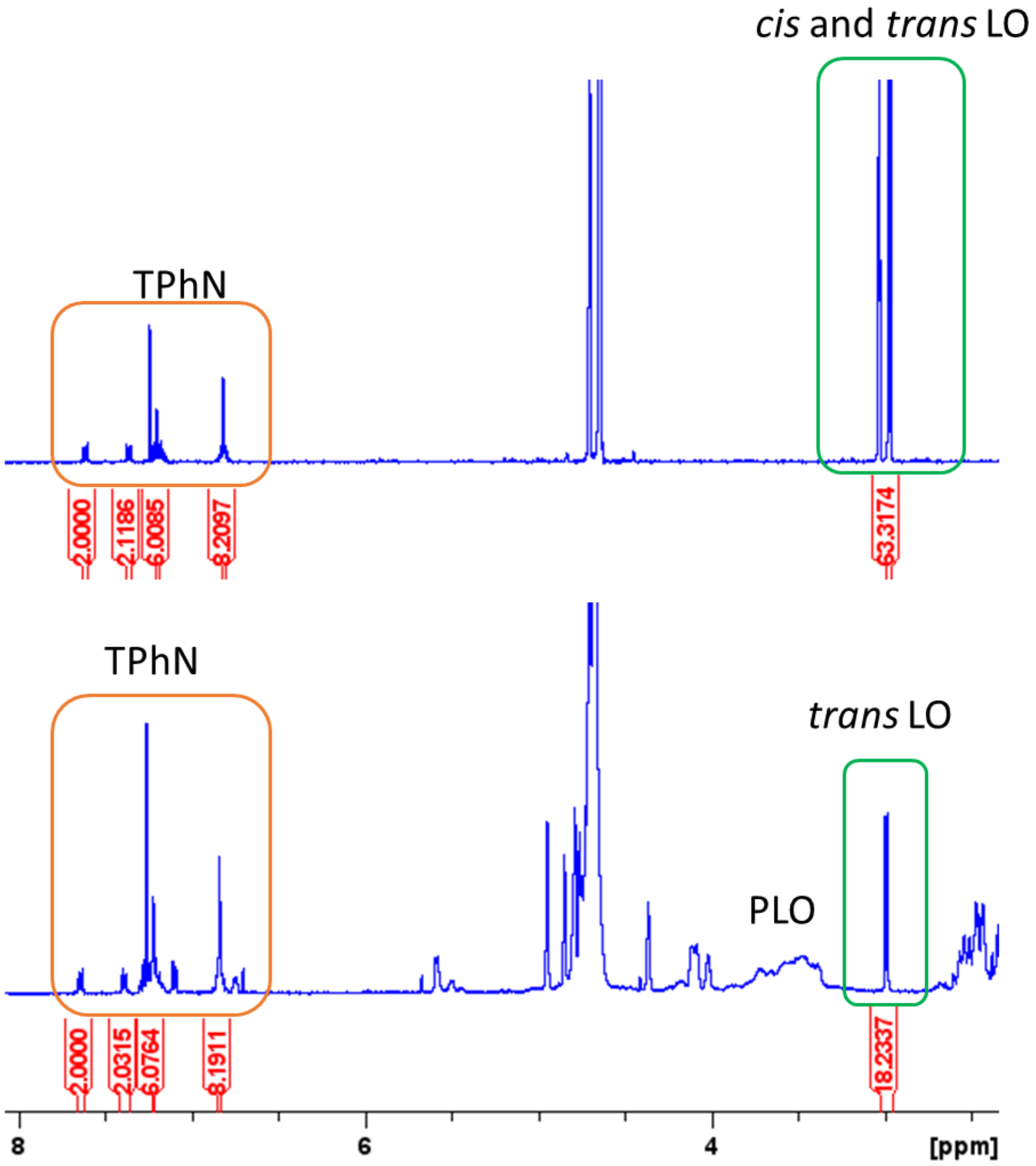

**Figure S6**. [1]H-NMR spectra in $CDCl_3$ of the reaction carried out with a [Al]:TPhN:LO ratio of 1:2:250 in bulk at 130 °C. (Top: before addition of catalyst; bottom: t=30 min)

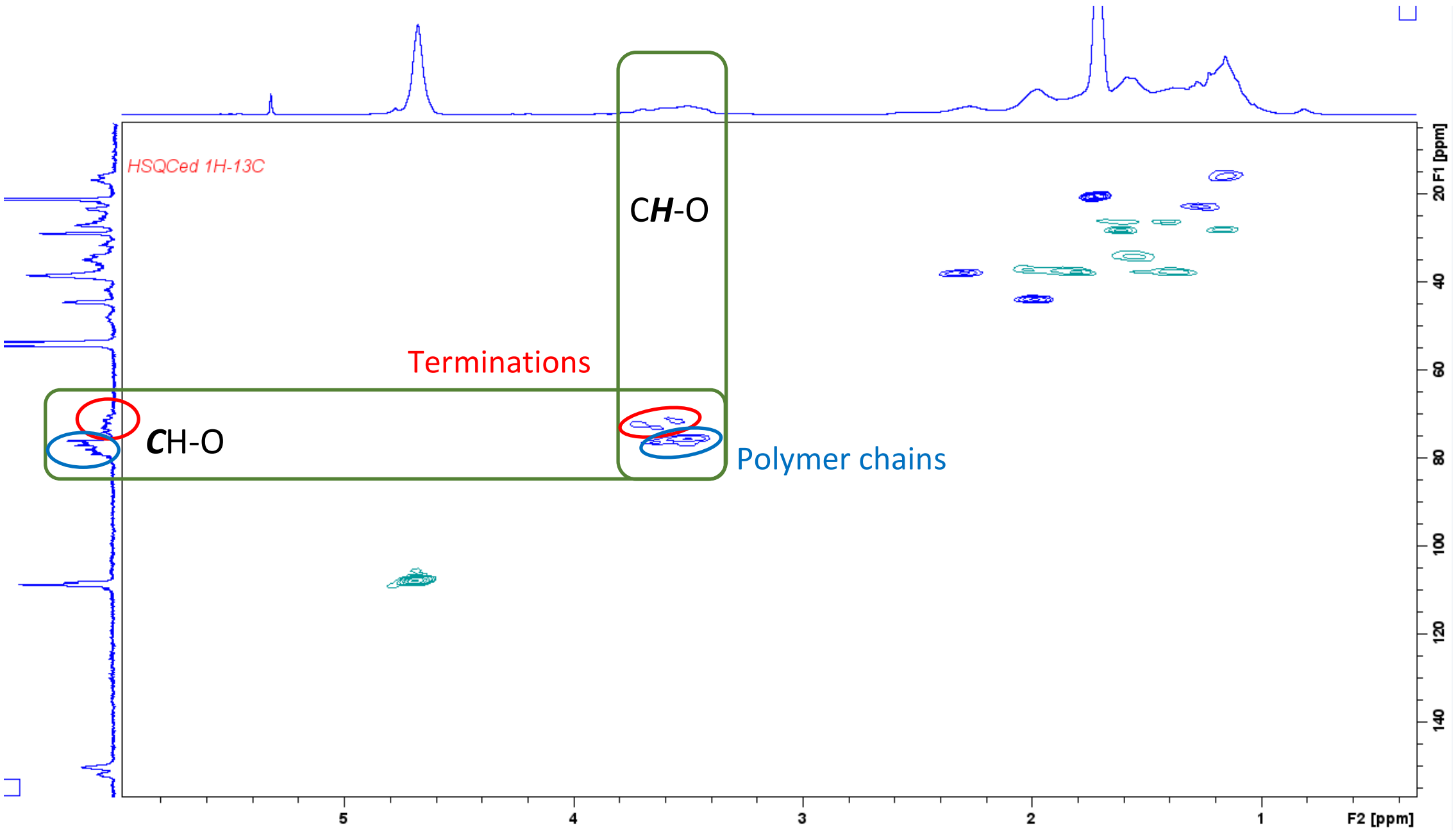

**Figure S7.** HSQCed $^{1}H$-$^{13}C$-NMR spectrum of PLO ($CD_2Cl_2$, 400 MHz, RT).

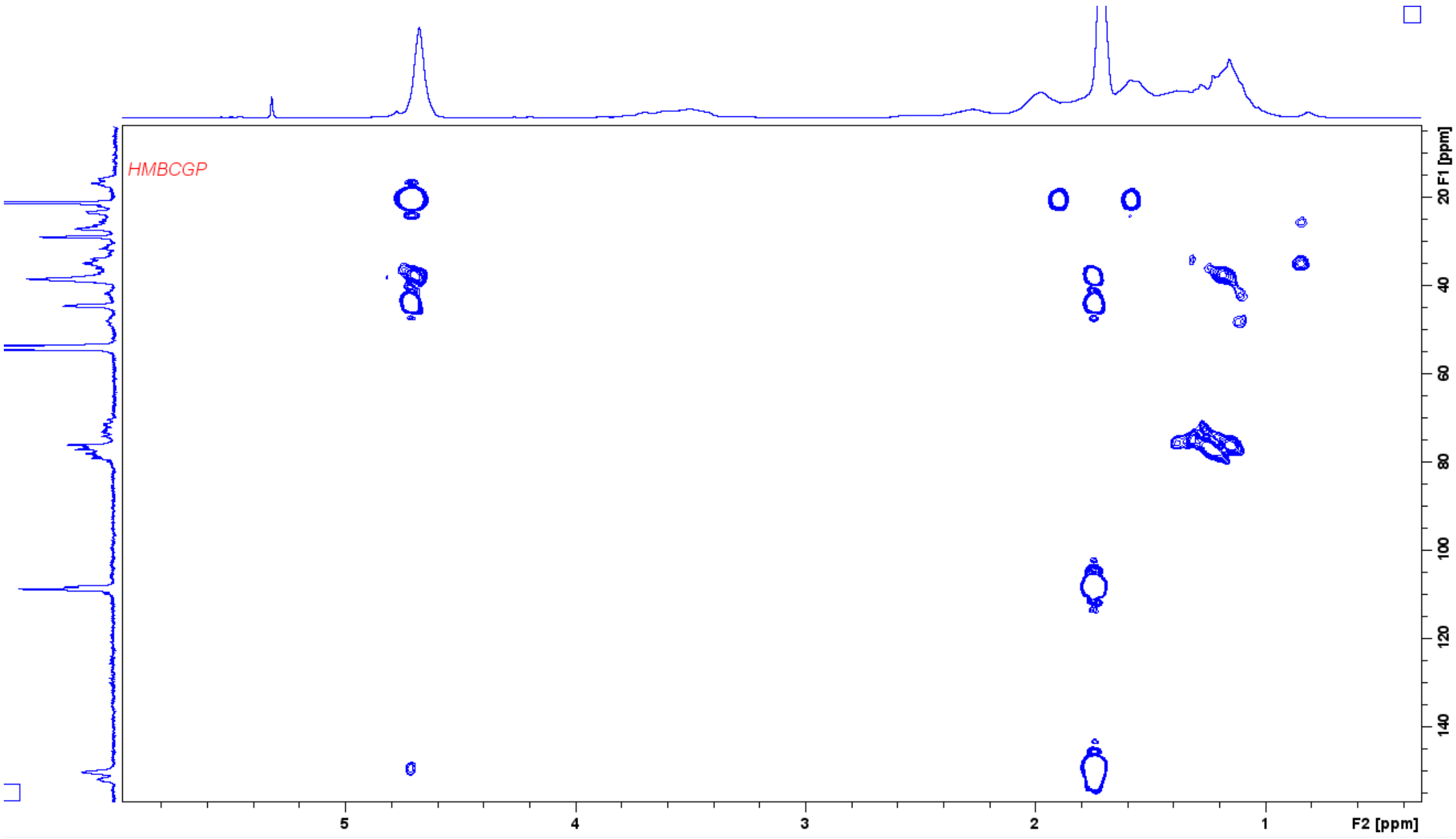

**Figure S8.** HMBC $^{1}H$-$^{13}C$-NMR spectrum of PLO ($CD_2Cl_2$, 400 MHz, RT)

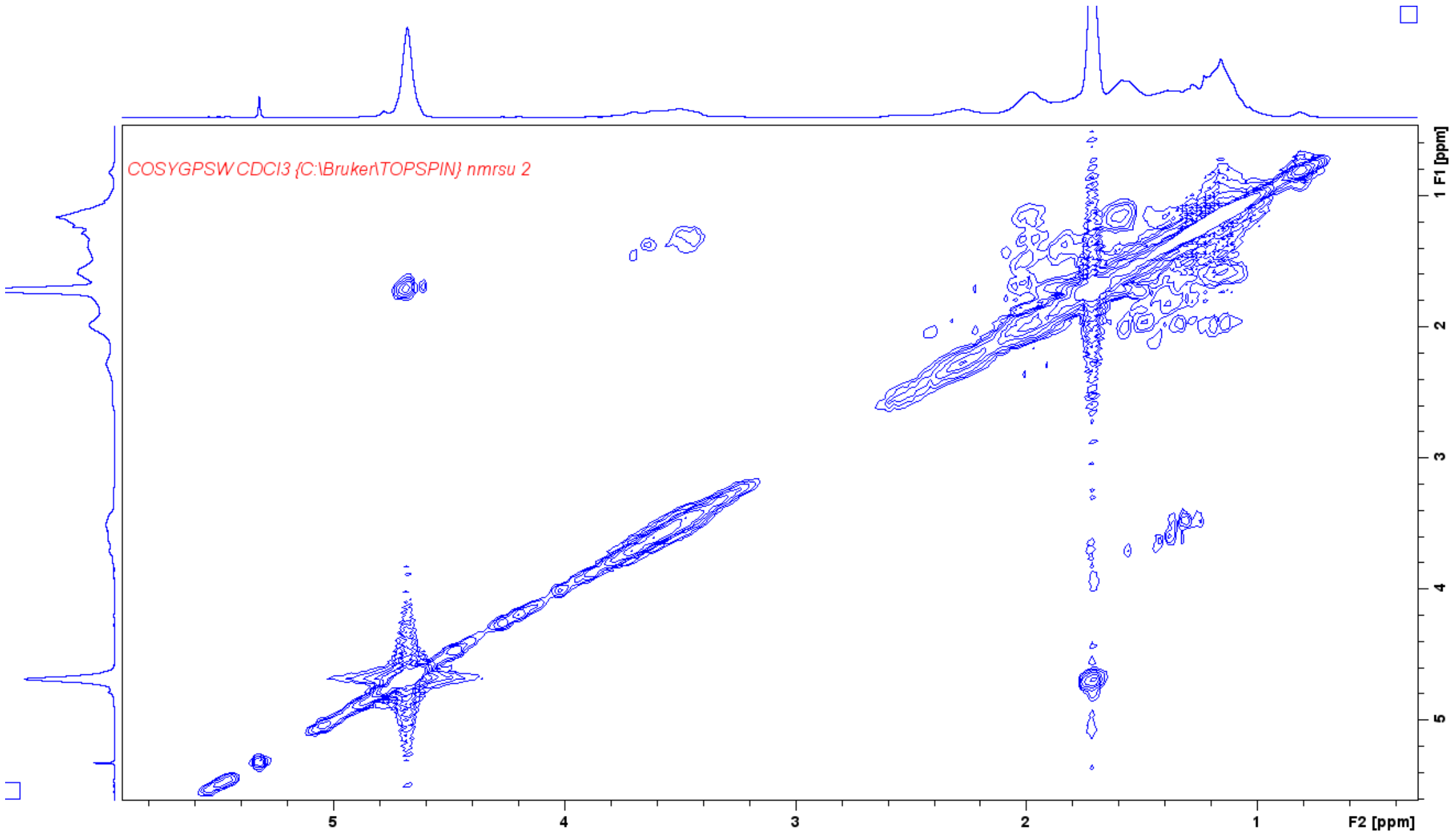

**Figure S9.** COSY [1]H-[1]H-NMR spectrum of PLO ($CD_2Cl_2$, 400 MHz, RT).

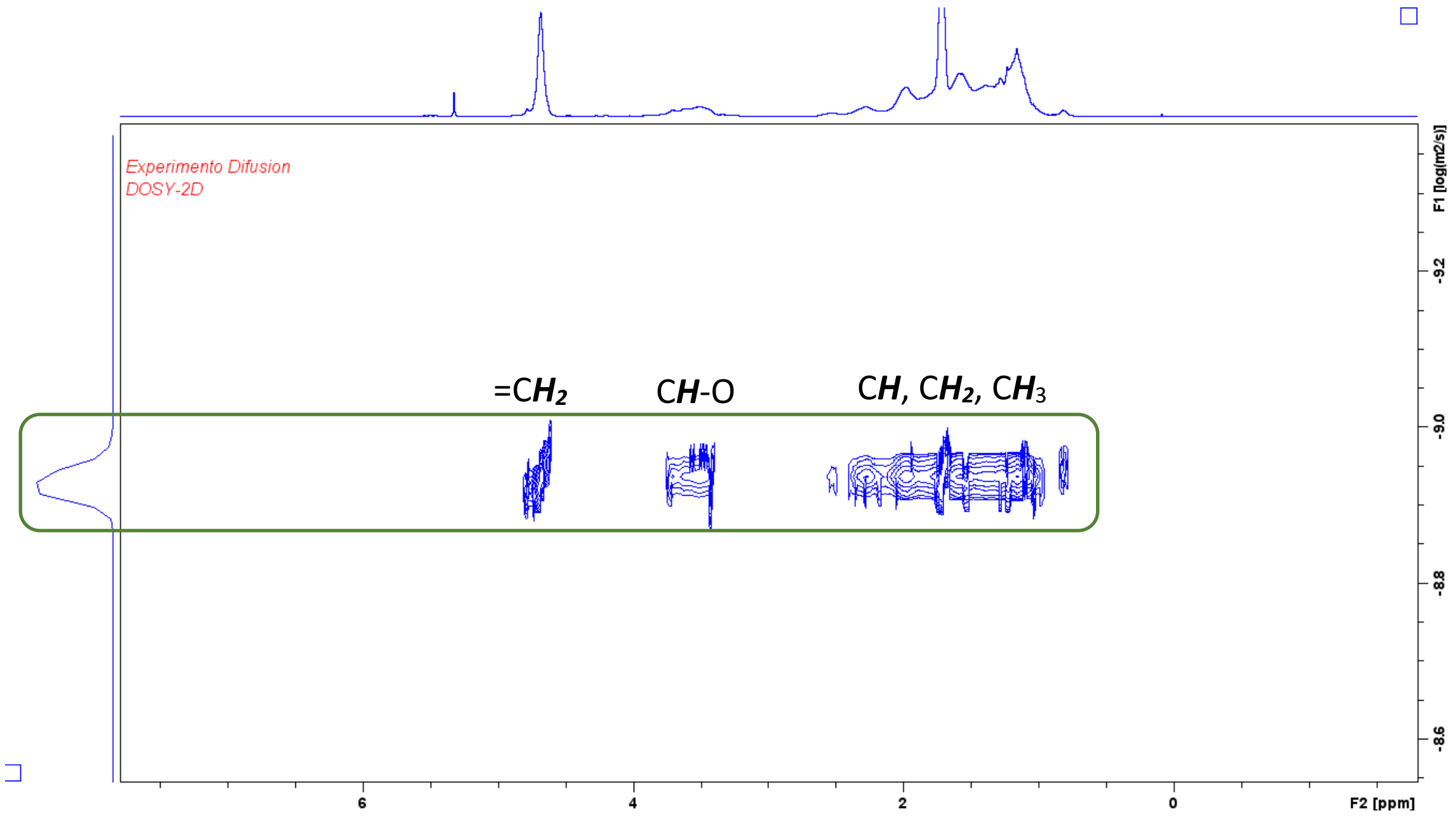

**Figure S10.** DOSY 2D NMR spectrum of PLO ($CD_2Cl_2$, 400 MHz, RT).

**Table S1.** Experiments of ROP of LO with catalysts 1 and 2.

| Ent. | [Al] | [Al]:[LO] | T (ºC) | %Conv. cis/trans-LO [a] | % Yield PLO |
|---|---|---|---|---|---|
| 1 | 1 | 1:100 | 130 | >99 / 70 | 28 |
| 2 | 1 | 1:100 | 25 | 94 / 83 | 54 |
| 3 | 1 | 1:250 | 130 | >99 / 70 | 28 |
| 4 | 1 | 1:250 | 25 | 80 / 67 | 60 |
| 5 | 2 | 1:100 | 130 | >99 / 70 | 28 |
| 6 | 2 | 1:100 | 25 | 91 / 78 | 60 |
| 7 | 2 | 1:250 | 130 | >99 / 70 | 28 |
| 8 | 2 | 1:250 | 25 | 77 / 62 | 30 |

[a] Determined by $^{1}$H-NMR spectroscopy.

## 3. Isothermal analysis under air atmosphere

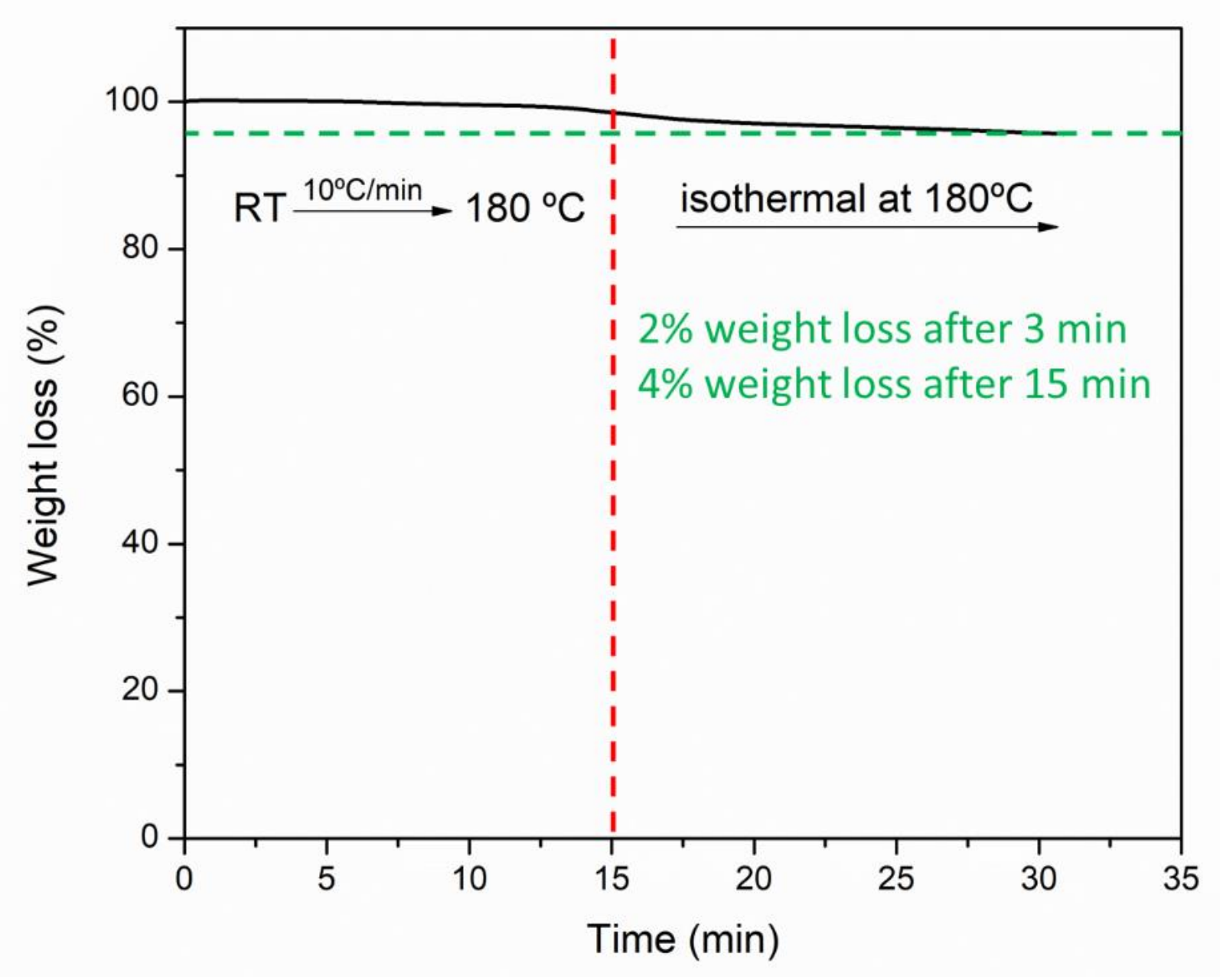

**Figure S11.** Isothermal TGA scan at 180ºC under air flow.